\documentclass[a4paper,twocolumn]{Gaia2004} 
\usepackage{times}      
\usepackage{epsfig}     
\usepackage{natbib}     

\title{Relativistic formulation and reference frame}

\author{Sergei A. Klioner}
\affil{Lohrmann Observatory, Dresden Technical University, Mommsenstr. 13, 01062 Dresden, Germany}

\bibpunct{(}{)}{;}{a}{}{,}  

\def\aj{AJ}%
\def\apjl{ApJ}%
\def\aap{A\&A}%
\def\mnras{MNRAS}%
\def\prd{Phys.~Rev.~D}%
\def\nat{Nature}%
\newcommand{\OO}[1]{{\cal O}(c^{-#1})}
\newcommand{\vecg}[1]{\mbox{\boldmath$#1$}}
\newcommand{\ve}[1]{\vecg{#1}}
\newcommand{\muas}[0]{\hbox{\rm $\mu$as}}

\def\newover#1{\mathop{\vtop{\ialign{##\crcr%
$\hfil\displaystyle{#1}\hfil$\crcr%
\noalign{\kern.1pt\nointerlineskip}%
\crcr\noalign{\kern.1pt}}}}\limits}

\def\newbot#1.#2{{\newover{#1}_{\scriptscriptstyle #2}}{}}
\def\nbot#1.#2{{\newover{#1}_{\scriptstyle #2}}{}}
\def\newtop#1.#2{{\newover{#1}^{\scriptscriptstyle #2}}{}}
\def\ntop#1.#2{{\newover{#1}^{\scriptstyle #2}}{}}

\begin{document}

\setlength{\parskip}{5 pt plus 1 pt minus 1 pt}


\maketitle

\begin{abstract}
After a short review of experimental foundations of metric theories of
gravity, the choice of general relativity as a theory to be used for
the routine modeling of Gaia observations is justified. General
principles of relativistic modeling of astronomical observations are
then sketched and compared to the corresponding Newtonian principles.
The fundamental reference system -- Barycentric Celestial Reference
System, which has been chosen to be the relativistic reference system
underlying the future Gaia reference frame is presented. Principal
relativistic effects in each constituent of a relativistic model of
astronomical observations are briefly elucidated. The structure of a
relativistic model of positional observations which can be used as a
standard relativistic model for Gaia is sketched. The physical
meaning of the Gaia reference frame is discussed. It is discussed also
how Gaia observations can be used to verify general relativity.
\end{abstract}

\section{Why relativity?}

Reduction scheme of positional observations in Newtonian physics is
rather simple. Absolute Euclidean space and absolute time of Newtonian
physics lead to the existence of global preferred coordinates: inertial
coordinates which are unique up to a constant shift of the origin of
the time coordinate, constant rotation of spatial axes and a shift
of the origin of spatial coordinates which is at most linear in time. Although
already in Newtonian physics one can introduce arbitrary coordinates
(e.g., some curvilinear coordinates), the inertial coordinates are
certainly preferred since the laws of physics look especially simple
when expressed in an inertial reference system. Moreover, observed
quantities (distances, directions, etc.) are directly related to those
global inertial coordinates.

 \begin{figure}[h]
  \begin{center}
    \leavevmode
\centerline{\epsfig{file=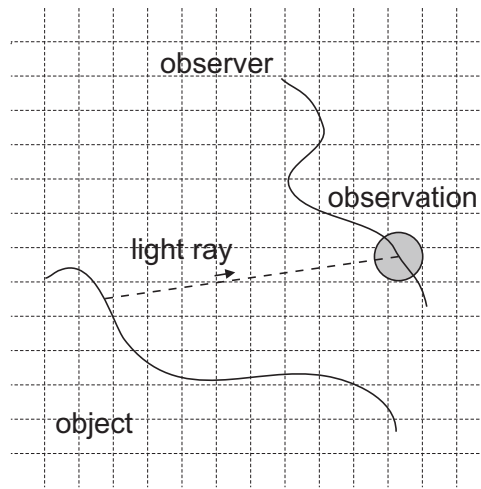,width=0.8\linewidth}}
   \end{center}
\caption{Four parts of an astronomical event from the point of view of Newtonian
physics: 1) motion of
the observed object; 2) motion of the observer; 3) trajectory of an
electromagnetic signal from the observed object to the observer which is
tacitly assumed to be a straight line in Newtonian astronomy; 4)
the process of observation responsible for Newtonian aberration.
The coordinate grid in the background symbolizes a global inertial
reference system.}
\label{Figure-astro-obs-Newton}
\end{figure}

Let us briefly consider the Newtonian scheme of reduction of astronomical
observations. Figure \ref{Figure-astro-obs-Newton} sketches
the four constituents of an astronomical observation from the point of
view of Newtonian physics: (1) motion of the
observed object, (2) motion of the observer, (3) propagation of an
electromagnetic signal from the object to the observer, and (4) the
process of observation. The last two parts can be formulated in a quite
simple way in Newtonian physics. It is normally tacitly assumed here
that the light rays are straight lines in some inertial coordinates. As
for ``the process of observation'', it is responsible for the appearance of
Newtonian aberration which reflects the difference in observed
directions to the source by a moving observer and by an observer at
rest relative to the chosen coordinates.

The goal of Newtonian reduction of astronomical observations is to
model (to predict) the results of observations performed by a
fictitious observer (normally situated at the origin of the chosen
reference system, e.g. at the barycenter of the solar system) at some
given moment of time. One attempts here to correct  for all the effects
in observations which are produced by the motion and the
position of the real observer (aberration and, e.g., parallax,
respectively) and by the motion of the object (proper motion and,
possibly, light travel time effects). The structure of a Newtonian
reduction scheme does not depend on the goal accuracy of reduction and
can be described as follows: (1) aberration, (2) parallax, (3) proper
motion and/or light travel time effects. For low accuracies when only
linear effects from aberration, parallax and proper motion are of
interest, one could apply the corresponding corrections in arbitrary
order. On the contrary, for higher accuracies the order of these
reductions is important. All parameters of the model, i.e.
the coordinates of the observer and the object as function of time, are
defined in the chosen inertial reference system. That is, the five
standard astrometric parameters of the object (right ascension
$\alpha$, declination $\delta$, parallax $\pi$, proper motion in right
ascension $\mu_\alpha$ and proper motion in declination $\mu_\delta$)
are also defined in the chosen reference system.

Rapid increase of observational accuracy of astronomical observations
has already made indispensable to use general relativity for modeling
of the observational data. For many kinds of observations the Newtonian
scheme sketched above fails to describe observational data with the
required accuracy. In many cases the deviations from the model are
several orders of magnitude larger than the accuracy of observations.
Examples are astrometric (geodetic) VLBI observations, lunar laser
ranging, radar ranging to the planets, experiments with high accuracy
clocks, GPS observations. It is also widely known and accepted that
the deviations can be eliminated by using Einstein's general theory of
relativity (instead of Newtonian physics) for the modeling of observations.

The accuracy of positional observations to be produced by Gaia is
expected to attain 2-3 \muas\ for the stars with magnitude $V<10$ mag and
10 \muas\ for the stars of $V=15$ mag. It is clear that not only the
largest relativistic effects, but also many additional subtle effects
should be taken into account to attain that accuracy. It is also quite
clear that relativistic effects cannot be considered as small
corrections to a Newtonian model as has been often done earlier when the
accuracy was not so high. The whole model should be formulated in a
language compatible with general relativity. In such a relativistic
framework many Newtonian concepts must be abandoned and the meaning of
astrometric parameters such as position, parallax and proper motion of
a star should be redefined.

\section{Experimental foundations of general relativity}

Einstein's general relativity is by no means the only possible theory
of gravity. However, it seems to be the simplest theory among the
theories successfully passing all available observational tests.  Let us
briefly review the experimental foundations of general relativity. A
detailed review of the modern experimental foundations of gravitational
physics can be found in \cite{Will:2001}.

\subsection{Einstein Equivalence Principle}

The basic principle of the theory is called Einstein Equivalence
Principle. This principle consists of the following three parts:

(1) The {\sl Weak Equivalence Principle} stating that the masses on the both
sides of the Newtonian gravitational law
\begin{displaymath}
m_{iner}\,\ddot r^i = -G\, m_{grav}\,M\,r^i\,/\,r^3
\end{displaymath}
\noindent
exactly coincide $m_{iner} \equiv m_{grav}$ for all bodies (actually,
this is equivalent to the claim that $G$ is a constant and its value is
independent of the choice of the bodies with which we measure it). The
Weak Equivalence Principle has been tested in many different
experiments with a precision of $|\delta m|/m < 4\cdot 10^{-13}$.

(2) {\sl Local Lorentz invariance} stating that the outcome of any local
non-gravitational experiment is independent of the velocity of the
freely-falling test laboratory (reference frame) where it is performed.
This is equivalent to the principal postulate of special relativity theory
which states that the light velocity in vacuum $c$ is constant in any
inertial reference system. This has been tested at a level of $\sim
10^{-21}$.

(3) {\sl Local positional invariance} which states that the outcome of any
local non-gravitational experiment is independent of where and when in
the universe it is performed. A part of local positional invariance can
be tested by measuring of the gravitational red shift (e.g., of the clock frequency)
\begin{displaymath}
\Delta\nu\,/\,\nu = (1+\alpha)\,c^{-2}\,\Delta U,
\end{displaymath}
\noindent
where $\alpha=0$ in general relativity. A number of different
experiments has proved that $|\alpha| < 2\cdot10^{-4}$. Another part of
local positional invariance (independence of ``position'' in time) can
be tested by looking for possible time-dependencies of fundamental
(non-gravitational) constants. Different kinds of experimental data
show tight constrains on possible time-dependence of the constants
(e.g., the fine structure constant should be constant at a precision of
$\sim 10^{-6}$ over Hubble time of 13 billion years).

\begin{table*}[t!]
  \caption{Most important post-Newtonian tests of general relativity}
  \label{Table-ppn-tests}
  \begin{center}
    \leavevmode
        \begin{tabular}[h]{llll}
      \hline \\[-5pt]
      observational data & relativistic effect      &  possible deviation     &  reference \\
      &       &  from general relativity     &   \\[+5pt]
      \hline \\[-5pt]
      VLBI  & differential Shapiro delay & $\pm0.0003$ & \citet{Eubanks:et:al:1997} \\
      HIPPARCOS & light deflection & $\pm0.003$ & \citet{Froeschle:Mignard:Arenou:1997} \\
      Viking radar ranging & Shapiro delay & $\pm0.002$ & \citet{Reasenberg:at:al:1979} \\
      Cassini radar ranging & Shapiro delay & $\pm0.000023$ & \citet{Bertotti:et:al:2003} \\
      planetary observations & perihelion advance & $\pm0.0002$ & \citet{Pitjeva:2001} \\
      Lunar laser ranging & Nordtvedt effect & $\pm0.001$ & \citet{Williams:et:al:1996} \\
      Lunar laser ranging & geodetic precession & $\pm0.001$ & \citet{Williams:et:al:1996} \\
      \hline \\
      \end{tabular}
  \end{center}
\end{table*}

\subsection{Testing metric theories of gravity}

One can argue that if the Einstein Equivalence Principle is valid the
gravity can be interpreted as an effect of curved spacetime. However,
the Einstein Equivalence Principle does not necessarily imply general
relativity. There exists a class of alternative theories of gravity
compatible with that Principle. These theories are called metric
theories of gravity. In order to test the principal observable effects of
metric theories of gravity a special scheme called Parametrized
Post-Newtonian (PPN) formalism has been proposed \citep{Will:1993}. The
scheme involves up to 10 numerical parameters which have different
values in different theories and which can be fitted from observations.
The most important PPN parameters are $\gamma$ and $\beta$. Results of
data processing with a PPN reduction model incolve a set of
constraints on the PPN parameters. Alternatively, the results can be
interpreted as boundaries on possible deviations from general
relativity. The most important experimental results in this second
interpretation are summarized in Table \ref{Table-ppn-tests}.

An additional test of general relativity is the search for possible
time-dependence of Newtonian gravitational constant $G$. This can be
done by looking for secular changes in semi-major axes of solar system
planets (especially, Mercury, Venus, Earth and Mars) as well as from
pulsar timing of double pulsars. The most stringent estimate here is
$|\dot G/G|<10^{-13}$ per year
\citep{Pitjeva:2001}. General relativity predicts that $G$ is
time-independent. One more argument in favor of general relativity is
the well-known indirect evidence for gravitational radiation in the double pulsar
timing data. Gravitational radiation is strong-field regime phenomenon
which is beyond the scope of the PPN formalism.

All this shows that one can be ``reasonably confident'' about the correctness of general
relativity, and that general relativity can be used as ``standard''
theory. Nevertheless, it is still very important to test general
relativity further. This will be discussed below in Section
\ref{Section-gaia-for-relativity}.

\section{Relativistic modeling of astronomical observations}

\begin{figure}[h]
  \begin{center}
    \leavevmode
\centerline{\epsfig{file=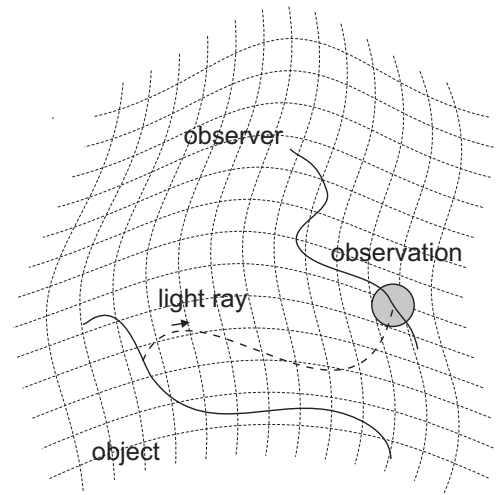,width=0.8\linewidth}}
   \end{center}
\caption{Four parts of an astronomical event from the point of view of
relativistic physics: 1) motion of the observed object; 2) motion of
the observer; 3) trajectory of an electromagnetic signal from the
observed object to the observer which represents a geodetic line (i.e. a
``curved'' line) in the chosen reference system; 4) the process of
observation. The grid of curved coordinates in the
background symbolizes the chosen relativistic reference system.}
\label{Figure-astro-obs-Einstein}
\end{figure}

Let us now outline general principles of relativistic modeling of
astronomical observations. It is interesting that in spite of a deep
conceptual difference between Newtonian physics and general
relativity, the structure of the reduction scheme changes, in
principle, only in one point: light rays are no longer straight lines
and should be carefully modeled. Figure \ref{Figure-astro-obs-Einstein}
shows the four constituents of an astronomical observation in the
relativistic framework. In curved spacetime there is no preferred
coordinates where the laws of physics would have substantial simpler
form than in other coordinates. Therefore, any reference system
covering the spacetime region under study can be used. Instead of
Newtonian inertial coordinates one has to choose some reference system
in curved spacetime which is sketched symbolically on Figure
\ref{Figure-astro-obs-Einstein} as a grid of curved coordinates.

\subsection{General scheme of relativistic modeling}

\begin{figure}[h]
  \begin{center}
    \leavevmode
\centerline{\epsfig{file=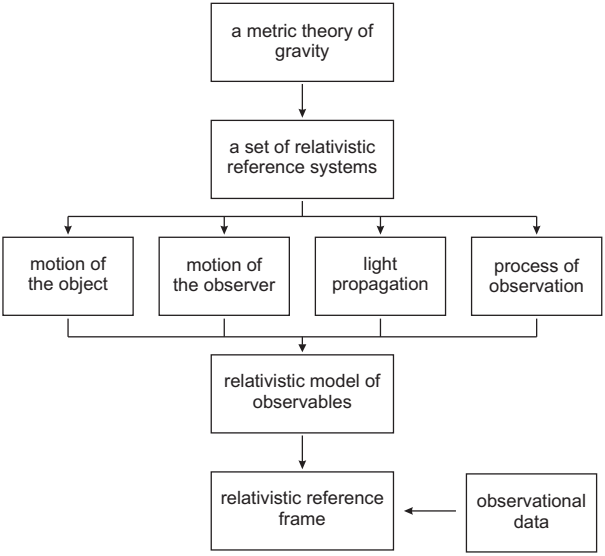,width=1.0\linewidth}}
   \end{center}
\caption{General principles of relativistic modeling of
astronomical observations (see text for further explanations).}
\label{Figure-scheme}
\end{figure}

General scheme of relativistic modeling is presented on Figure
\ref{Figure-scheme}. Starting from general theory of relativity, any
other metric theory of gravity or the PPN formalism one should define
at least one relativistic 4-dimensional reference system covering the
region of space-time where all the processes constituting particular
kind of astronomical observations are located. Each of four
constituents of an astronomical observation should be modeled in the
relativistic framework. The equations of motion of both the observed
object and the observer relative to the chosen reference system should
be derived and a method to solve these equations should be found. The
equations of light propagation relative to the chosen reference system
should be derived and a way to solve them should be found. The
equations of motion of the object and the observer and the equations of
light propagation enable one to compute positions and velocities of the
object, observer and the photon (light ray) with respect to that
particular reference system at a given moment of the coordinate time,
provided that the positions and velocities at
some initial epoch are known. However, the positions and velocities
calculated in this way obviously depend on the reference system, that
is on the preferences of the person who writes down the equations. On the
other hand, the results of observations cannot depend on the choice of
the reference system. Therefore, it is clear that one more step of the
modeling is needed: a relativistic description of the process of
observation. This part of the model allows one to compute a
coordinate-independent theoretical prediction of observables starting
from the coordinate-dependent quantities mentioned above.

These four components can now be combined into relativistic models of
observables. The models give an expression for relevant observables as
a function of a set of parameters. These parameters can then be fitted
to observational data using some kind of parameter estimation scheme.
The sets of certain estimated
parameters appearing in the relativistic models of observables
represent astronomical reference frames (see Section \ref{Section-Gaia-RF}).
It is important to understand at this point that the relativistic
models contain some parameters which are defined only in the chosen
reference system(s) and are thus coordinate-dependent.
For example, position and velocity of the observed object are clearly
coordinate-dependent.

\subsection{The Barycentric Celestial Reference System}

From the physical point of view any reference system covering the
region of space-time under consideration can be used to describe
physical phenomena within that region. In this sense we are free to
choose the reference system to be used to model the observations.
However, reference systems, in which mathematical description of
physical laws is in one sense or another simpler than in some other
reference systems, are more convenient for practical calculations.
Therefore, one can use the freedom to choose the reference system to
make the parametrization as convenient and reasonable as possible.

Two Working Groups on relativity in astrometry, celestial mechanics and
metrology established 1997 by the International Astronomical Union
(IAU) and Bureau International des Poids and Mesure (BIPM) have come to
the conclusion that the most convenient relativistic reference system
for the applications in astrometry, solar system dynamics, and time
keeping and dissemination is defined by the following metric tensor
\citep{Soffel:et:al:2003}:
\begin{eqnarray}
\label{BCRS_metric}
g_{00} &=& -1 + {2w \over c^2} - {2w^2 \over c^4} + \OO5, \nonumber\\
g_{0i} &=& -{4 \over c^3} w^i+\OO5, \label{bary_metric} \nonumber\\
g_{ij} &=& \delta_{ij}\left(1 + {2 \over c^2}w \right) + \OO4 \,
\end{eqnarray}
\nonumber
with the post-Newtonian potentials $w$ and $w^i$ defined by
\begin{eqnarray}\label{fielda}
w(t,\ve{x}) &=& G \int d^3 x' \,
{\sigma(t, \ve{x}') \over \vert \ve{x} - \ve{x}' \vert}
\nonumber\\
&&
 + {1 \over 2c^2} G {\partial^2 \over \partial t^2}
\int d^3 x' \, \sigma(t,\ve{x}') \vert \ve{x} - \ve{x}' \vert \, ,
\\
\label{fieldb}
w^i(t,\ve{x}) &=& G \int d^3 x' {\sigma^i (t,\ve{x}') \over
\vert\ve{x} - \ve{x}' \vert } \, ,
\end{eqnarray}
\noindent
$\sigma$ and $\sigma^i$ being related to the components
of the energy-momentum tensor $T^{\alpha\beta}$:
\begin{equation} \label{sigma}
\sigma = {1\over c^2}\left(T^{00} + T^{ss}\right), \quad
\sigma^i = {1\over c}\ T^{0i}\, .
\end{equation}
\noindent
The origin of spatial coordinates of this reference system is chosen to
coincide with the barycenter of the solar system. The reference system
defined in this way is called Barycentric Celestial Reference System
(BCRS). The BCRS has been explicitly recommended by the IAU for the
modeling of high accuracy astronomical observations
\citep{IAU:2001,Rickman:2001,Soffel:et:al:2003}. For the moment the
BCRS is a post-Newtonian reference system with higher order terms
(post-post-Newtonian terms, etc.) neglected in the metric tensor
(\ref{BCRS_metric}). The reason for that is that the post-Newtonian
approximation is sufficient to model any observations in forseeable
future (including microarcsecond astrometry as long as the observations
are made further than about one degree from the Sun).
Post-post-Newtonian terms can be added to the metric tensor as soon as
they are necessary for some applications. The word ``celestial'' in the
name of BCRS is used to underline that the BCRS do not rotate with the
Earth and that remote sources do not move relative to the
BCRS in some averaged sense. The second reference system deined by the same IAU resolutions
\citep{Rickman:2001} is the Geocentric Celestial Reference System
(GCRS). This reference system is only marginally important for Gaia
(mostly for modeling of orbit tracking data and relating the Gaia
onboard clock to TAI \citep{Klioner:2003}) and will not be discussed here.
The PPN version of the BCRS valid for certain class of metric
theories of gravity can be found in \cite{Klioner:Soffel:2000} and
\citet{Will:1993}.

The BCRS will be also used for the modeling of Gaia observations. This
is a reference system underlying the resulting Gaia catalogue (see
Section \ref{Section-Gaia-RF} below). The coordinate time of the BCRS
is called Barycentric Coordinate Time (TCB). The TCB will be used to
parametrize the Gaia catalogue.

\subsection{Motion of the objects and the observer}

Typically, for objects situated in the solar
system (asteroids, planets, space vehicles) the equations of motion are
ordinary differential equations of second order and numerical
integration with suitable initial or boundary conditions can be used to
solve them. For objects outside of the solar system one use often
simple models like uniform and rectilinear motion in space or more
complicated ones, e.g., for binary stars. In any case one should understand
that in the relativistic framework all these ad hoc models give positions
and velocities of observed objects in the chosen relativistic
reference system.

The principal relativistic effects in the translational motion
of bodies in the solar system (including Gaia satellite, asteroids, etc.)
are contained in the so-called Einstein-Infeld-Hoffmann (EIH)
equations of motion of $N$ gravitating bodies, whose gravitational fields
can be described by their masses $M_A$ only:
\begin{eqnarray}
\ddot{\ve{x}}_A &=& - \sum_{B\neq A} G\,M_B\,
{\ve{x}_A-\ve{x}_B\over |\ve{x}_A-\ve{x}_B|^3}
\nonumber\\
&&+{1\over c^2}\,{\bf F}_{pN}(M_B,\ve{x}_B,\dot{\ve{x}}_B)+{\cal O}(c^{-4}).
\end{eqnarray}
\noindent
The Newtonian part of these equations (shown explicitly above) follows
from the term of order $c^{-2}$ in $g_{00}$. The relativistic terms
require all
other terms in the BCRS metric tensor specified above. Various parts of
these equations represent:  (1) relativistic perihelion advance
($\sim$43 {\hbox{$^{\prime\prime}$}}/cty for Mercury, $\sim$10
{\hbox{$^{\prime\prime}$}}/cty for Icarus, etc.); (2) geodetic
precession ($\sim$2 {\hbox{$^{\prime\prime}$}}/cty for Lunar orbit);
(3) various periodic relativistic effects (important mostly for LLR and
binary pulsar timing observations). Further effects not contained in
the EIH equations are the effects due to rotation of the bodies
(Lense-Thirring or gravitomagnetic effects) and those due to
non-sphericity of the gravitating bodies. These additional effects are
marginal for the current accuracy of LLR and SLR, but negligible for
Gaia. In case of Gaia satellite one should use a slightly simplified
version of the EIH equations since the influence of the mass of the satellite
on the motion of other gravitating bodies can be neglected.

The BCRS metric tensor allows one also to derive the equations of
rotational motion of an extended body. These equations will not be
discussed here, since they are not important for Gaia.

\subsection{Light propagation}

In any metric theory of gravity the equations of light propagation
coincide with the equations of geodetic lines in the chosen reference
system. The latter are ordinary differential equations of second order.
These equations could also be solved by numerical integrations, but
normally one prefer to use some approximate analytical solutions. Only
in some special (normally, highly symmetrical) cases like Schwarzschild
metric exact analytical solutions are known. Anyway, an appropriate way
to solve the equations of light propagation should be found.

The structure of the BCRS equations of light propagation can be written as follows
\begin{eqnarray}
\ve{x}(t)=\ve{x}_0&+&c\,\ve{\sigma}\,(t-t_0)+c^{-2}\,{\bf S}_{pN}(t)
\nonumber\\
&+&
c^{-3}\,{\bf S}_{1.5pN}(t),
\end{eqnarray}
\noindent
where
$\ve{x}_0$ and $\ve{\sigma}$ are the parameters of Newtonian straight line,
${\bf S}_{pN}$ are the post-Newtonian terms, and ${\bf
S}_{1.5pN}$ are the additional effects induced by the motion of
gravitating matter (i.e., by translational and rotational motion of
gravitating bodies). The terms of order of $c^{-2}$ in both $g_{00}$
and $g_{ij}$ are required to derive ${\bf S}_{pN}(t)$, and the terms
$c^{-3}$ in $g_{0i}$ are needed for ${\bf S}_{1.5pN}$. The next order
effects, the so-called post-post-Newtonian effects, would require terms
of order of $c^{-4}$ in both $g_{00}$ and $g_{ij}$ (the $c^{-4}$ terms
in $g_{ij}$ are not in the current definition of the BCRS metric
tensor). The principle observable effects in the light propagation are
(1) the gravitational light deflection (amounting to $1.75
{\hbox{$^{\prime\prime}$}}$ for a light ray grazing the Sun) and (2)
the gravitational signal retardation (the Shapiro effect; this effect
amounts to $\sim240$ $\mu$s for the radar ranging of Venus in upper
conjunction).

\subsection{Conversion to observables: proper direction}

As mentioned above the conversion the coordinate-dependent quantities
into coordinate-independent observables is an important part of
relativistic modeling. From the mathematically point of view the
coordinate-independent quantities are scalars. Special mathematical
techniques are known to perform the suitable conversion in each particular case.
One of the most important application of this conversion procedure is a
conversion of the coordinate direction $\ve{n}$ into the source into
the corresponding observable direction $\ve{s}$. The observable
direction is often called ``proper direction'' in gravitational
physics. Proper direction is a direction relative to the proper
reference frame of the observer
(see Section \ref{Section-Gaia-RF} about the difference of the
concept of ``reference frame'' in astronomy and gravitational physics).
A proper reference frame is a
mathematical model of an ideal clock and three orthogonal rigid rods
which the observer uses to measure time intervals, distances and
directions in his vicinity. In special theory of relativity
the proper reference frame of an observer is related to some
inertial reference system by a
Lorentz transformation. It is therefore, sufficient to use Lorentz
transformations to convert $\ve{n}$ into $\ve{s}$.
The parameter of the Lorentz transformation in this case coincides with
the velocity of the observer relative to the chosen reference system.
In general relativity
it is also sufficient to
use Lorentz transformations, but the parameter $\ve{\nu}$
of the transformations
should be related to the BCRS velocity of the observer as
\begin{equation}
\label{nu-xdot}
\ve{\nu}=\dot{\ve{x}}_o\,\left(1+{2\over c^2}w(t,\ve{x}_o)\right)+\OO4,
\end{equation}
\noindent
where $\ve{x}_o$ and $\dot{\ve{x}}_o$ are the BCRS positions and velocity of
the observer, respectively.
A detailed discussion of this conversion and comparison of
different approaches can be found in \citet{Klioner:2003b}.
The relativistic terms in (\ref{nu-xdot}) are derived from
the $c^{-2}$ terms in $g_{00}$ and $g_{ij}$ of the BCRS metric tensor.
The difference between $\ve{n}$ and $\ve{s}$ can be called
relativistic aberration. The difference between the Newtonian
aberration and the relativistic one may amount to
several milliarcsecond for Gaia observations.

\subsection{Conversion to observables: proper time}

Another important case is the conversion of intervals of the coordinate
time $t$ into the corresponding intervals of the proper time $\tau$ of
the observer. The general form of this conversion reads
\begin{equation}
d\tau\,/\, dt=1+c^{-2}\,A_{pN}+c^{-4}\,A_{ppN}+{\cal O}(c^{-5}),
\end{equation}
\noindent
where $A_{pN}$ and $A_{ppN}$ are the post-Newtonian and
post-post-Newtonian terms, respectively. Explicit form of these two
functions depends on the metric tensor: in order to compute for
$A_{pN}$ the $c^{-2}$ terms in $g_{00}$ are needed, while
the $c^{-4}$ terms  in $g_{00}$, the $c^{-3}$ terms in
$g_{0i}$, and the $c^{-2}$ ones in $g_{ij}$ are required to compute
$A_{ppN}$. Typically in the Solar system and in particular
for Gaia onboard clocks $|c^{-2}\,A_{pN}|\sim 10^{-8}$
and $|c^{-4}\,A_{ppN}|\sim 10^{-16}$.

\section{Relativity for Gaia}

Now, having all these theoretical tools one can formulate the
relativistic model for Gaia. The relativistic model for Gaia is well
documented \citep{Klioner:2003}, so that we just outline the overall
structure of the model here.

\subsection{Structure of the standard relativistic model}

 \begin{figure}[t]
  \begin{center}
    \leavevmode
\centerline{\epsfig{file=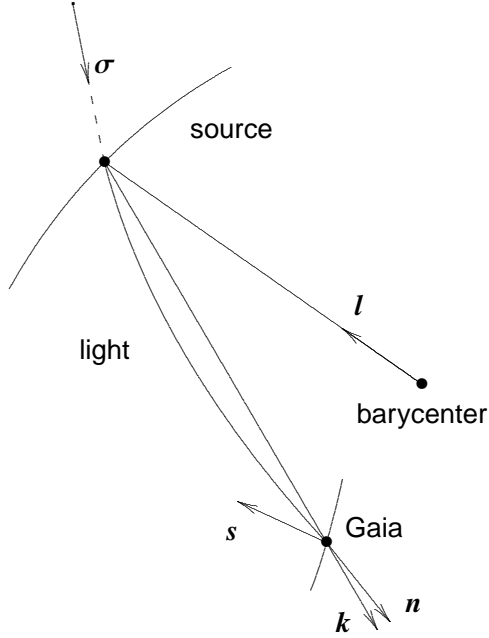,width=0.8\linewidth}}
   \end{center}
\caption{Five principal vectors used in the model (see text for explanations).}
\label{Figure-5-vectors}
\end{figure}

The model consists essentially in subsequent transformations between 5 following
vectors (Figure \ref{Figure-5-vectors}):

a) $\ve{s}$ is the unit observed direction (the word ``unit'' means here
and below that the formally Euclidean scalar product
$\ve{s}\,\cdot\,\ve{s}= s^i\,s^i$\ is equal to unity),

b) $\ve{n}$ is the
unit vector tangential to the light ray at the moment of observation,

c) $\vecg{\sigma}$ is the unit vector tangential to the light ray at
$t=-\infty$,

d) $\ve{k}$ is the unit coordinate vector from the source to
the observer,

e) $\ve{l}$ is the unit vector from the barycenter of the
Solar system to the source.

Note that the last four vectors should be interpreted as sets of three
numbers characterizing the position of the source with respect to the
BCRS. Vector $\ve{s}$ represents components of the observed
direction relative to the local proper reference system of the
satellite. All these vectors would change their numerical values if
some other relativistic reference system is used instead of the BCRS.
The model consists then in a
sequence of transformations between these vectors as shown on
Figure \ref{Figure-transformations}. The physical meaning of each
transformation can be summarized as follows (the numbering here coincides
with the numbering on Figure \ref{Figure-transformations}):

\begin{figure*}[t!]
{\large
\hskip2cm \hbox to 35mm{remote sources:\hfill}\quad
$
\ve{s}\ \newtop \longleftrightarrow.{(1)}\
\ve{n}\ \newtop \longleftrightarrow.{(2)}\
\ve{\sigma}\ \newtop \longleftrightarrow.{(3)}\
\ve{k}\ \newtop \longleftrightarrow.{(4)}\
\ve{l}, \pi \ \newtop \longleftrightarrow.{(5)}\
\ve{l}(t_0), \pi(t_0), \ve{\mu}(t_0), \dots
$
\vskip 3mm

\hskip2cm\hbox to 35mm{solar system objects:}\quad
$
\ve{s}\ \newtop \longleftrightarrow.{(1)}\
\ve{n}\ \newtop \longleftrightarrow.{(2,3)}\
\ve{k}\ \newtop \longleftrightarrow.{(6)}\
{\rm orbit}
$
}
\caption{Transformation sequences (see text for explanations).}
\label{Figure-transformations}
\end{figure*}

(1) aberration (effects vanishing together with the barycentric
velocity of the observer): this step converts the observed direction to
the source $\ve{s}$ into the unit BCRS coordinate velocity of the light ray
$\ve{n}$ at the point of observation;

(2) gravitational light deflection for the source at infinity:  this step
converts $\ve{n}$ into the unit direction of propagation $\ve{\sigma}$
of the light ray infinitely far from the solar system at $t\to-\infty$;

(3) coupling of finite distance to the source and the
gravitational light deflection in the gravitational field of the solar
system: this step converts $\ve{\sigma}$ into a unit BCRS coordinate
direction $\ve{k}$ going from the source to the observer;

(4) parallax: this step converts $\ve{k}$ into a unit BCRS direction
$\ve{l}$ going from the barycenter of the Solar system to the source;

(5) proper motion, etc: this step provides a reasonable parametrization
of the time dependence of $\ve{l}$ (and, possibly, of the parallax
$\pi$) caused by the motion of the source relative to the barycenter of
the Solar system;

(6) orbit determination process.

These transformations have already been discussed in full detail
\citep{Klioner:2003,Klioner:Peip:2003,Klioner:2003b}. Let us only
mention the following. The most complicated part of the model is the
light deflection model where the effects of (1) monopole fields of all
major solar system bodies, (2) quadrupole fields of the giant planets,
and (3) gravitomagnetic fields due to translational motion of all major
bodies should be taken into account in order to attain the accuracy of
1 \muas. Moreover, each body with a mean density $\rho$ and radius
$R\ge (\rho/ 1\ {\rm g/cm}^3)^{-1/2}\times 650$ km produces a light
deflection of at least 1 \muas. Therefore, a few tens of minor bodies
(mainly, satellites of the giant planets) should also be taken into account
in certain rare cases \citep{Klioner:2003}. The parametrization of time
dependence of $\ve{l}$ in the relativistic framework looks exactly the
same as in the Newtonian case. The only difference is that all vectors and
parameters here (parallax, proper motion, etc.) are coordinate
quantities defined in the BCRS.

\subsection{Implementation of the model}

An ANSI C code has been written to implement the relativistic model in
its full complexity \citep{Klioner:Blankenburg:2003,Klioner:2003c}. The model has
been implemented in two modes: predictor mode and corrector mode. {\sl
Predictor} mode implements the standard way of astrometric reductions
when the observed direction to the source is predicted starting from
some a priori catalogue parameters (coordinates, proper motion,
parallax, etc.) of that source. The catalogue is supposed to be
improved later by fitting the parameters to the whole set of data. {\sl
Corrector} mode implements the reductions in the opposite direction, that is,
the momentary barycentric direction to the source is restored from the
observed direction as good as possible. The code does not contain any
attempt to restore parallaxes and proper motions, or orbits of the
sources: the model can only be applied separately for each individual
observation.

Both modes are implemented both for solar system objects and for remote
sources situated outside of the solar system. The principal difference between
remote sources and solar system sources lies in the treatment of the
light propagation: an initial value solution of the corresponding
differential equations is used in the former case, while in the latter
case two point boundary value problem should be solved. Although
analytical approximations are used in both cases, a rather
time-consuming numerical inversion process is used for solar system
sources in the predictor mode in order to attain the goal accuracy of 1
\muas. The corresponding refinement of the model aimed at direct
analytical solution for this case is underway
\citep{Klioner:Blankenburg:2004}.

For remote sources the corrector and predictor modes being implemented
independently of each other must give exactly the opposite
transformations. This was used to massively test the implementation
\citep{Klioner:Blankenburg:2003}. The situation is different for solar
system objects. Here for the corrector mode, it is statistically better
to calculate the gravitational light deflection as if the body were a
remote source, even if it is known a priori that the source is a solar
system body (but it is not known how far the body is, otherwise at
least a preliminary orbit is known and one should better use the
predictor mode).

The implementation was conceived to be as flexible as possible.
Internal parameters allow one to select the type of arithmetic to be
used (to test possible numerical instabilities), to change easily any
of the physical, mathematical and astronomical constants used in the
model (including switching between several available planetary
ephemerides), to switch on and off each individual effect. Both
predictor and corrector mode routines have a goal accuracy parameter,
which is used together with some a priori criteria to decide which effects should
be computed in each particular case. The latter feature allows one to
speed up the calculations substantially if a lower accuracy is
sufficient (e.g. the source brightness information can be used to meet
the Gaia observational accuracy for fainter objects).

The implementation has been used in massive numerical tests aimed at
identifying possible inconsistencies or numerical instabilities as well
as points of critical numerical performance. The implementation was
used also to test another simplified model implementation used in the
GDAAS \citep{Anglada:2004}. The implementation of the full model will be
further supported and optimized.

\subsection{Beyond the standard relativistic model}

The model described above is constructed under assumption that the
solar system is isolated. This means that any influence of
gravitational fields generated outside of the solar system are ignored
in the model. For the majority of the sources the external field can
indeed be fully neglected, but there are a number of cases when the
external gravitational fields produce observable effects. Several
authors have discussed these additional effects in detail (see, e.g.,
\citet{Klioner:2003} and \citet{Kopeikin:Gwinn:2000}). Let us briefly
list here the main effects of this kind:

(1) Gravitational light deflection caused by the masses situated outside of the solar system:
(a) weak microlensing on the stars of the Galaxy \citep{Belokurov:Evans:2002},
(b) lensing on gravitational waves (both primordial ones and those
from compact sources),
(c) lensing of the companions of edge-on binary systems.

(2) Cosmological effects.

(3) More complicated models for the motions of observed objects
in the BCRS are necessary for the case of binary stars, etc.

Note that all these effects can be easily taken into account by a simple
additive extension of the standard model since at the required
accuracy the external gravitational fields can be linearly superimposed
on the solar system gravitational field. The only exception could be
the effects of cosmological background, but a preliminary study by
\citet{Klioner:Soffel:2004} shows that even here the coupling of the
local solar system fields and the external ones can be neglected.

\section{Gaia reference frame}
\label{Section-Gaia-RF}

It is important to remember that all astrometric parameters of
sources obtained from Gaia observations will be defined in the BCRS
coordinates: positions, proper motions, parallaxes, radial velocities,
orbits of minor planets, binaries, etc. All these parameters will
represent the Gaia reference frame, which is a materialization of the
BCRS. The Gaia reference frame is, so to say, a model of
the universe in the BCRS. Thus, the goal of astrometry in the
relativistic framework is not to find ``the'' barycentric inertial
reference system, which is unique in Newtonian formulation, but to find
a materialization of some chosen relativistic reference system.

Let us note here that the meaning of words ``reference system'' and
``reference frame'' in relativistic astronomy is different from the
meaning normally used in gravitational physics. {\sl Reference system}
is a purely mathematical construction (a chart) giving ``names'' to
space-time events. A {\sl reference frame} is, in contrast, some
materialization (realization) of a reference system. In astronomy the
materialization is normally given in a form of a catalogue (or
ephemeris) containing positions of some celestial objects relative to
the selected reference system. Any astronomical reference frame (a
catalogue, an ephemeris, etc) is defined only through the reference
system(s) used to construct physical models of observations.

\section{Gaia for Relativity}
\label{Section-gaia-for-relativity}

Using general relativity for the standard reduction model does not mean
that Gaia data should not be used to test general relativity itself. On
the contrary, testing relativity is one of the exciting goals of the mission.
Gaia will certainly deliver an estimate of the PPN parameter $\gamma$,
appearing mainly in the magnitude of the light deflection effects, with
an unprecedented accuracy
of $\sim5\times10^{-7}$. However, it is by no means the
only way Gaia will improve our knowledge of gravitational physics.
Gravitational light deflection could be tested in a much more profound
way with the Gaia data Gaia. It will be certainly possible to
look for terms with totally different dependence on the angular
distance to the deflecting body. In this sense one should be able to get
first experimental estimates of higher-order effects. Also the PPN
parameter $\beta$, appearing in the equations of motion of solar system
bodies, will be determined with an accuracy of $\sim 10^{-4}$
\citep{Hestrofer:Berthier:2004} which is comparable with the current
accuracy from the planetary fits \citep{Pitjeva:2001}. A simultaneous
fit with the planetary data could further improve the accuracy. Special
data processing of observations close to Jupiter and Saturn (which
should be dropped from the global absolute solution since the positions
of those planets are not known with an accuracy necessary to predict
the light deflection at the level of 1 \muas) will allow to test subtle
relativistic effects caused by translational motion of the planets and
by their quadrupole gravitational fields (see
\cite{Crosta:Mignard:2004} for a preliminary study of the second of
these possibilities). A number of cosmological tests (upper estimates of
parallaxes of quasars, apparent proper motions of quasars,
possible traces of low-frequency gravitational waves in those proper
motions, direct measurement of the acceleration of the solar system
relative to quasars, etc.) will also have a big impact on our
knowledge. A lot of work is still necessary to understand how to use
the full potential of the huge amount of observation data Gaia will deliver
to us in the most efficient and useful way.

\section*{Acknowledgments}

The author is grateful to F.Mignard, M.Soffel and the members of the
Gaia Relativity and Reference Frame Working Group for numerous
stimulating discussions.

\end{document}